\documentclass[preprint]{emulateapj}

\slugcomment{Accepted for Publication in  the Astrophysical Journal}
\shortauthors{Zaritsky et al.}
\shorttitle{Lopsidedness in S$^4$G Galaxies}

\begin{document}
\title{On the Origin of Lopsidedness in Galaxies as Determined from the Spitzer Survey of Stellar Structure in Galaxies (S$^4$G) }
  
\author{Dennis Zaritsky\altaffilmark{1}, Heikki Salo\altaffilmark{2}, Eija Laurikainen\altaffilmark{2,3}, 
 Debra Elmegreen\altaffilmark{4}, E. Athanassoula\altaffilmark{5}, Albert Bosma\altaffilmark{5}, S\'ebastien Comer\'on\altaffilmark{2}, Santiago Erroz-Ferrer\altaffilmark{6,7}, Bruce Elmegreen\altaffilmark{8}, Dimitri A. Gadotti\altaffilmark{9}, Armando Gil de Paz\altaffilmark{10}, Joannah L. Hinz\altaffilmark{1,11}, Luis C. Ho\altaffilmark{12}, Benne W. Holwerda\altaffilmark{13}, Taehyun Kim\altaffilmark{9,14,15}, Johan H. Knapen\altaffilmark{6,7}, Jarkko Laine\altaffilmark{2}, Seppo Laine\altaffilmark{16}, Barry F. Madore\altaffilmark{12}, Sharon Meidt\altaffilmark{17},
Karin Menendez-Delmestre\altaffilmark{18}, Trisha Mizusawa\altaffilmark{13,16,19}, Juan Carlos Mu\~noz-Mateos\altaffilmark{9,14}, Michael W. Regan\altaffilmark{20}, Mark Seibert\altaffilmark{12},  Kartik Sheth\altaffilmark{14}}

\altaffiltext{1}{Steward Observatory, University of Arizona, 933 North Cherry Avenue, Tucson, AZ 85721, USA; dennis.zaritsky@gmail.com}
\altaffiltext{2}{Astronomy Division, Department of Physics, P.O. Box 3000, FI-90014 University of Oulu, Finland}
\altaffiltext{3}{Finnish Centre of Astronomy with ESO (FINCA), University of Turku, V\"ais\"al\"antie 20, FI-21500, Piikki\"o, Finland}
\altaffiltext{4}{Vassar College, Dept. of Physics and Astronomy, Poughkeepsie, NY 12604, USA}
\altaffiltext{5}{Aix Marseille Universit\'e, CNRS, LAM (Laboratoire
d'Astrophysique de Marseille) UMR 7326, 13388, Marseille, France}
\altaffiltext{6}{Instituto de Astrof\'isica de Canarias,
V\'{i}a L\'{a}cteas 38205 La Laguna, Spain}
\altaffiltext{7}{Departamento de Astrof\'isica, Universidad de La Laguna, 38206, La Laguna,  Spain}
\altaffiltext{8}{IBM T. J. Watson Research Center, 1101 Kitchawan Road, Yorktown Heights, NY 10598, USA}
\altaffiltext{9}{European Southern Observatory, Casilla 19001, Santaigo 19, Chile}
\altaffiltext{10}{Departamento de Astrof\'isica, Universidad Complutense de Madrid, Madrid, Spain }
\altaffiltext{11}{MMT Observatory, P.O. Box 210065, Tucson, AZ 85721, USA}
\altaffiltext{12}{The Observatories of the Carnegie Institution for Science, 813 Santa Barbara Street, Pasadena, CA 91101, USA}
\altaffiltext{13}{ESA Fellow, European Space Agency (ESTEC), Keplerlaan 1, 2200 AG Noordwijk, The Netherlands}
\altaffiltext{14}{National Radio Astronomy Observatory/ NAASC, 520 Edgemont Road, Charlottesville, VA 22903, USA }
\altaffiltext{15}{Astronomy Program, Department of Physics and Astronomy, Seoul National University, Seoul 151-742, Korea}
\altaffiltext{16}{Spitzer Science Center, 1200 East California Boulevard, Pasadena, CA 91125, USA}
\altaffiltext{17}{Max-Planck-Institut f\"ur Astronomie, K\"onigst\"uhl 17 D-69117, Heidelberg, Germany}
\altaffiltext{18}{Observatorio do Valongo, Universidade Federal de Rio de Janeiro, Ladeira Pedro Ant™nio, 43, Sa\'ude CEP 20080-090, Rio de Janeiro-RJ-Brazil }
\altaffiltext{19}{California Institute of Technology, 1200 East California Boulevard, Pasadena, CA 91125, USA}
\altaffiltext{20}{Space Telescope Science Institute, 3700 San Martin Drive, Baltimore, MD 21218, USA }


\begin{abstract} 
We study the $m=1$ distortions (lopsidedness) in the stellar components of 167 nearby galaxies that span a wide range of morphologies and luminosities. We confirm the previous findings of 1) a high incidence of lopsidedness in the stellar distributions, 2) increasing lopsidedness as a function of radius out to at least 3.5 exponential scale lengths, and 3) greater lopsidedness, over these radii, for galaxies of later type and lower surface brightness. Additionally, 
the magnitude of the lopsidedness 1) correlates with the character of the spiral arms (stronger arm patterns occur in galaxies with less lopsidedness), 2) is not correlated with the presence or absence of a bar, or the strength of the bar when one is present, 3) is inversely correlated to the stellar mass fraction, $f_*$, within one radial scale length, and 4) correlates directly with $f_*$ measured within the radial range over which we measure lopsidedness. We interpret these findings to mean that
lopsidedness is a generic feature of galaxies and does not, generally, depend on a rare event, such as a direct accretion of a satellite galaxy onto the disk of the parent galaxy.  While lopsidedness may be caused by several phenomena, moderate lopsidedness ($\langle A_1 \rangle_i + \langle A_1 \rangle_o)/2 < 0.3$) is likely to reflect halo asymmetries to which the disk responds or a gravitationally self-generated mode . We hypothesize that the magnitude of the stellar response depends both on how centrally concentrated the stars are with respect to the dark matter and whether there are enough stars in the region of the lopsidedness that self-gravity is dynamically important.  
\end{abstract}


\section{Introduction}
\label{sec:intro}
When studying physical systems, it is often useful to understand how they respond when disturbed. The responses to various perturbations can highlight the internal structure and the physical processes involved. Signatures of perturbations often involve departures from symmetry. For galaxies, these departures vary from the dramatic, large tidal tails, bridges, and shells \citep{arp}, to common features such as bars and spiral arms. The study of the latter has a long and rich history, is at the core of the morphological classification of galaxies \citep[see][]{sandage}, and continues to be a morphological touchstone to the current day,  even for the sample of galaxies that will be discussed here \citep{buta,elmegreen}. 

With the advent of digital detectors, large surveys, and computing power, and the desire to make results as reproducible as possible, quantitative measures of morphological features have been developed and implemented. For example, the strength of bars and spiral arms can be quantified in terms of the amplitude of the $m=2$ Fourier decomposition of the azimuthal surface brightness distribution \citep{considere,grosbol,elmegreen89}. In comparison to the extensive study of $m=2$ modes, and the role these have in phenomena as varied as star formation \citep{roberts,elmegreen86} and the fueling of the central supermassive black holes \citep{schwarz,noguchi}, the $m=1$ asymmetries are comparatively neglected.

The $m=1$ features, first noted systematically in the H {\small I} distributions of galaxies \citep{baldwin}, are a common feature of stellar disks as well \citep{block}, with somewhere between 20\% and 50\% of galaxies exhibiting what is considered to be strong lopsidedness \citep{rix,bournaud,reichard08}, where ``strong" has commonly been defined to mean that the surface brightness asymmetry from one side of the galaxy to other is greater than 20\% of the mean surface brightness at that radius, although other threshold choices are valid \citep[for example,][advocate using 10\%]{bournaud, jog}. Although such features are generically referred to as ``lopsidedness", $m=1$ distortions are only truly a measurement of lopsidedness if the phase of the component is constant, or nearly so, with radius. This distinction has been noted in some discussions of the $m=1$ decompositions \citep{li}, but is sometimes neglected. In general, when considered, the conclusion is that the distortions are indeed characteristic of lopsidedness \citep{jog97, bournaud, angiras, eymerenb}. Despite the ubiquity of $m=1$ distortions, how they arise remains 
unresolved, even though various mechanisms have been suggested \citep[see][for a review of the field]{jog}. As usual, the scenarios can be divided into a ``nature" category, for example one in which the lopsidedness arises from an asymmetric dark matter dominated potential, and a ``nurture" category, for example one in which the lopsidedness arises directly from interactions or mergers with another galaxy. Disentangling the scenarios, if at all possible, requires more information than has been available, but is critical in enabling us to use lopsidedness to learn about either fundamental properties of galaxies or to reconstruct critical aspects of their evolution.

In general, there are several concerns that observational studies of disk asymmetries face. First, one wants to measure the asymmetry in the underlying stellar mass, not light. This desire tends to push the studies to redder bandpasses to avoid the strong influence of recent star formation episodes. However, as noted since the early near-IR studies \cite[for example][]{rix,rhoads} these concerns are not entirely answered by going as far red as possible because young and intermediate age evolved stars will also bias the spatial distribution of infrared flux. Recent work has attempted to reach a quantitative understanding of this contribution and has demonstrated that this issue is a non-negligible source of uncertainty \citep{melbourne,meidt,eskew}. Second, one wants to mitigate extinction, which certainly affects the measurement of $m=2$ features because dust can trace spiral structure and bars, but is perhaps less of a factor in the measurement of $m=1$ features. Both of these concerns help drive studies to the red, at least to $I$ band and often into the near-infrared. Third, one wants a large sample with which to search for correlations with other physical parameters. This goal has motivated the study of large samples in  visible light \citep{reichard08}, although there are other limitations, such as spatial resolution and surface brightness limits, that temper one's ability to construct ever larger samples, particularly in the near-infrared. The S$^4$G sample is the largest infrared sample of nearby galaxies to date and reaches significantly lower surface brightnesses than the previous near-infrared samples as well as most visible-light samples. 

In this paper, we present a study of the behavior of lopsidedness in the stellar distribution of galaxies, reaching conclusions regarding the distribution of lopsidedness among the galaxy population, the radial behavior of lopsidedness, and the correlations between lopsidedness and Hubble type and surface brightness that agree with those reached in the studies cited above. We then proceed beyond these results by examining correlations between the nature of $m=2$ modes and lopsidedness, and the correlation of lopsidedness with more detailed physical quantities, such as the stellar surface density and the stellar mass fraction, in an attempt to identify the dominant reason for lopsidedness in galaxies. We will conclude that lopsidedness is a generic feature of galaxies, rather than a common but externally triggered phenomenon, and suggest that the strength of the distortion is connected to physical characteristics of the stellar and dark matter distributions.

\section{The Data and Measurements}
\label{sec:data}

The parent sample for this study is the S$^4$G sample \citep{sheth}, which now consists of 2,352 galaxies. This sample is believed to be representative within the local volume, although additional selection criteria or requirements imposed by us, such as the existence of an H {\small I} redshift, and the surface brightness limitations of the existing catalogs from which the sample was selected, may introduce bias. Given the degree to which we are nearly complete to a magnitude limit of $m_B = 15.5$ with respect to existing catalogs, we expect that the sample is representative \citep[for more information about selection see][]{sheth}. We observed these galaxies using the {\sl Spitzer Space Telescope} \citep{werner} and its Infrared Array Camera \citep[IRAC,][]{irac} at 3.6 and 4.5$\mu$m and reduced the data as described by \cite{regan} and \cite{jc}. In this study we use only the 3.6$\mu$m data. All of the data are now publicly available (http://irsa.ipac.caltech.edu/data/SPITZER/S4G/). 

\begin{figure}[htbp]
\plotone{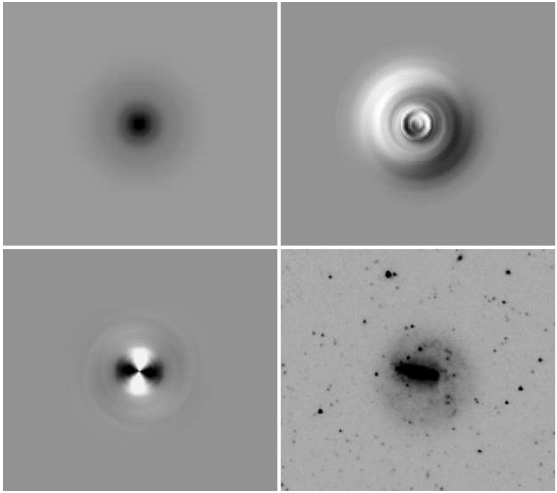}
\caption{Decomposition of $m=$ 0, 1, and 2 Fourier modes for NGC 3906. The lower right panel shows the galaxy image at 3.6$\mu$m. The scaling values are different in each panel to enable the reader to see the detailed structure. The absolute values of the $m=0$ mode are larger than those of the $m=1$ and 2 modes at all radii. The image has an angular scale of $\sim$ 4 arcmin on a side.}
\label{fig:decompose}
\end{figure}

The asymmetry analysis that we pursue is based on the azimuthal decomposition of the luminosity in circular annuli. For most such treatments, the images are deprojected to account for the inclination of the galaxy to the line of sight. While this issue is key for the measurement of the $m=2$ mode, which would be artificially inflated by inclination if left uncorrected, it is unimportant for the $m=1$ mode, at least as long as the inclination is moderate. Instead of deprojecting the image, we select nearly face-on (inclination $\le$ 30$^\circ$) galaxies to study, thereby avoiding uncertainties introduced by uncertainties in the measured inclination and position angle. We do correct the measured $m=2$ amplitudes for inclination, but these corrections are small. Selecting face-on galaxies also provides other benefits, such as diminishing the effect of internal extinction and enabling a less ambiguous definition of the galaxy center. We repeat the Fourier decomposition at each radius, where the radius is incremented by 2 pixels (1.5 arcsec) for each subsequent annulus. Before calculating the Fourier terms,
we apply the foreground stellar masks described by \cite{jc} and \cite{salo} and interpolate across the masked regions. The underlying galaxy luminosity, $I(r,\theta)$, is described as
\begin{equation}
I(r,\theta) = A_0 + \sum_{m=1}^\infty A_m(r) \cos(m(\theta-\theta_m(r)),
\end{equation}
where we find the best fitting $A_m$, for $0 \le m \le 4$, and $\theta_m$, for $0 < m \le 4$, and both $A_m$ and $\theta_m$ are functions of radius.
This technique, which has been used before \citep[see][for an example]{salo10}, is applied to all of the galaxies that have inclinations $\le$ 30$^\circ$ and T-type $> -5$. To be specific, $A_0$ measures the mean surface brightness as a function of radius, the amplitudes $A_m$ measure the strength of the $m$th component, where $m = 1,2,3,$ and $4$ in our decomposition, and the values of $\theta_m$ measure the corresponding phase angle of that component as a function of radius.
To compare among galaxies of different surface brightnesses, we express the strength of the $m$th component relative to the 0th component, $A_m/A_0$. The phases are arbitrarily referenced to the image axes and so only differences in phase have any physical meaning. We detect lopsidedness when the $m=1$ amplitude is large and the phase is roughly constant. We show an example of the results of the Fourier decomposition in Figure \ref{fig:decompose}.

Inclinations, for those galaxies having dynamically cold, thin disks, are typically determined by inverting the observed axis ratio under the assumption that disks are intrinsically round. In practice, this process involves selecting an isophote, typically, as we have also chosen, an outer one, for which one measures the major to minor axis ratio. We reviewed each image to ensure that we were measuring the disk rather than an extended halo component, that the surface brightness profile was regular at the chosen radius, and that the uncertainties in the surface brightness were still modest. In certain types of investigations, such as those attempting to constrain the intrinsic shape of disks (such as \cite{rix} or \cite{ryden}) this approach to determining inclination is clearly inappropriate. Instead, those studies must determine inclinations using a shape-independent approach, for example using the ratio of the H {\small I} line width, which is inclination dependent, to the expected width calculated from the Tully-Fisher relation and the galaxy's magnitude. Here, because we use inclination only as a mild selection criterion, demanding that the galaxy be less than 30$^\circ$ from face-on, and are not seeking to study the intrinsic disk shape, we do not expect inclination to play a major role and we adopt the determination based on the axis ratio. However, investigating the intrinsic shape of stellar disks remains an interesting question, and attempts to do that in the future with these data should not use the inclinations derived from the S$^4$G Pipeline 4 analysis \citep{sheth,salo}.

A few technical details merit some further discussion. As noted above, for several reasons we limit our sample to galaxies that are nearly face-on, with inclinations $\le 30^\circ$.
The exact value of the limiting inclination is subjective, and is a compromise between the inclusion of some projection effects and the retention of a sufficiently large sample. Because of the large size of the original sample, even a fairly strict inclination cut results in a statistically robust sample. For a randomly oriented sample of galaxies, an inclination cut of 30$^\circ$ results in the inclusion of only 14\% of the original sample. However, significantly increasing this fraction results in inclination cuts that we consider to be too permissive given our preference for selecting face-on systems. For example, doubling the fraction would require increasing the inclination cut to 46$^\circ$.
Second,  the galaxy center is as defined in the P4 stage of the S$^4$G
reductions and the method will be explained in detail by \cite{salo}.
Briefly, we visually identify the galaxy center to provide an initial estimate and a more precise center is determined algorithmically.
The automated procedure works well for brighter galaxies, but not for
some late-type systems. For those galaxies where the method fails, the center was
estimated visually. The important conceptual point is that the center is defined to be a local maximum in the surface brightness and not representative of what one might estimate as the center on the basis of an outer isophote. Although the identification of the center is one of the most obvious potential sources of systematic error in the determination of lopsidedness, in practice it does not affect the results. Any misplacement of the center in an axisymmetric galaxy will cause $A_1$ to increase sharply as $r \rightarrow 0$, rather than decrease as observed (see below), and will result in a decreasing $A_1$ as a function of radius, opposite to the observed trend. For both of these reasons, centering is not considered to be an important source of uncertainty in this discussion.

\begin{deluxetable*}{lrrrrrrcccccrr}
\tabletypesize{\tiny}
\tablecaption{The Sample}
\tablehead{
\colhead{Name}&
\colhead{T-}&
\colhead{D}&
\colhead{$M_{3.6}$}&
\colhead{$R_S$}&
\colhead{$\chi^2$}&
\colhead{Arm}&
\colhead{Bar}&
\colhead{$\langle A_1 \rangle_i$} &
\colhead{$\langle A_2 \rangle_i$}&
\colhead{$\langle A_1 \rangle_o$}&
\colhead{$\langle A_2 \rangle_o$}&
\colhead{$\langle \theta_1 \rangle_i$} &
\colhead{$\langle \theta_1 \rangle_o$} 
\\
&
&
\colhead{[Mpc]}&
\colhead{[AB]}&
\colhead{[pixels]}
\\
}
\startdata
ESO026-001 & 5.9 & 38.5 & $-$19.5 & 19.1 & 0.73 & 1 & 1 &0.086 $\pm$ 0.013 & 0.110 $\pm$ 0.014 & 0.057 $\pm$ 0.012 & 0.230 $\pm$ 0.013 &66 $\pm$ 18&$-$25 $\pm$ 20\\
ESO234-049 & 4.0 & 37.7 & $-$19.7 & 18.8 & 0.53 & 0 & 0 & 0.564 $\pm$ 0.011 & 0.106 $\pm$ 0.017 & 0.572 $\pm$ 0.015 & 0.340 $\pm$ 0.030 &133 $\pm$ 5&131 $\pm$ 4\\
ESO287-037 & 8.4 & 37.8 & $-$19.2 & 21.3 & 0.54 & 0 & 1 & 0.161 $\pm$ 0.019 & 0.318 $\pm$ 0.015 & 0.138 $\pm$ 0.016 & 0.134 $\pm$ 0.028 &$-$155 $\pm$ 15&$-$239 $\pm$ 33\\
ESO341-032 & 8.9 & 40.5 & $-$19.3 & 18.3 & 0.42 & 0 & 0 & 0.204 $\pm$ 0.030 & 0.090 $\pm$ 0.016 & 0.479 $\pm$ 0.044 & 0.070 $\pm$ 0.007 &157 $\pm$ 11&229 $\pm$ 8\\
ESO345-046 & 7.0 & 30.5 & $-$18.9 & 31.1 & 0.47 & 0 & 0 &0.208 $\pm$ 0.020 & 0.134 $\pm$ 0.019 & 0.273 $\pm$ 0.019 & 0.244 $\pm$ 0.026 &$-$337 $\pm$ 7&$-$464 $\pm$ 42\\
ESO359-031 & 7.8 & 58.7 & $-$19.0 & 14.4 & 0.63 & 0 & 0 & 0.320 $\pm$ 0.030 & 0.118 $\pm$ 0.018 & 0.636 $\pm$ 0.053 & 0.209 $\pm$ 0.031 &350 $\pm$ 17&379 $\pm$ 42\\
ESO407-018 & 9.8 &  1.3 & $-$11.0 & 28.1 & 0.47 & 0 & 1 & 0.245 $\pm$ 0.026 & 0.151 $\pm$ 0.031 & 0.222 $\pm$ 0.061 & 0.082 $\pm$ 0.011 &224 $\pm$ 6& 204 $\pm$ 177\\
\enddata
\tablenote{First seven entries for example purposes. The table is truncated in this rendering. Full electronic version will be available with published paper.}
\label{tab:data}
\end{deluxetable*}

The sample satisfying these criteria contains 186 galaxies. Most of the reduction in sample size from 2,352 to 186 is due to the inclination criteria. 
However, our final sample is smaller than one would expect if orientation were the only criteria (that sample would be expected to have 329 galaxies), which suggests that some additional hidden selection is occurring. We speculate that a surface brightness selection bias in both the catalog magnitudes and H {\small I} measurements is in part resulting in the over-representation of highly inclined systems in the S$^4$G sample. Furthermore, there is a bias that results from the application of internal extinction corrections that we applied, based on literature values of the inclination. These inclination estimates are often poor for faint galaxies. Noise in the inclination measurement is likely to result in 
the inclusion of highly inclined systems for which the correction is highly sensitive to inclination and can easily be overestimated, enabling the galaxy to satisfy the magnitude limit and enter the S$^4$G. 
Such a bias does not affect our study, other than in reducing the fraction of suitable systems for further study.

We obtain supplementary data from existing databases. Morphologies expressed as T-types are from the amalgamation provided by Hyperleda \citep{hyperleda}. While those may be somewhat irregular, with poorly defined uncertainties, we use them only as a rough sorting criteria and they are included in Table \ref{tab:data}. We adopt distances provided by NED, using their full local flow modeling (Virgo + Great Attractor + Shapley) and their default cosmological parameters, which are consistent with the standard $\Lambda$ cosmology.

To compare the properties of asymmetric features among galaxies, we define a radial range over which to calculate them because previous studies have shown them to be radius dependent \citep[for example, $m=1$ modes tend to increase in strength with radius,][]{rix}. We adopt an approach of the type used by \cite{zaritsky} who calculated the strength of the first Fourier component, $\langle A_1 \rangle$, as the average of $A_1/A_0$ between 1.5 to 2.5 disk scale lengths, $R_S$ (see below for how we calculate $R_S$). Averaging over a radial interval helps reduce the noise in these determinations and adopting a radial range that scales with galaxy size defines a fiducial that can be used across all galaxies. Here we expand on this approach by calculating the average amplitude of both $A_1/A_0$ and $A_2/A_0$, $\langle A_2 \rangle$, and do so not only over the radial interval between 1.5 and 2.5 $R_S$, but also over a second that spans 2.5 to 3.5 $R_S$. We distinguish results for the two radial ranges with the subscript $i$ for the inner of the two annuli and $o$ for the outer,  for example using $\langle A_1 \rangle_i$ to denote the average of $A_1/A_0$ over the inner radial range. The deep data from S$^4$G allow us to explore the behavior at larger radii than in most previous studies \citep[see][for an exception]{eymerenb}, but we do not extend beyond 3.5$R_S$ because we start to lose many galaxies from the sample due to the field-of-view. Outer disk features in S$^4$G galaxies are discussed by \cite{seppo}. We limit the inner radius to 1.5$R_S$ because interior to 1.5$R_S$ other components, bars and bulges, become important and also theoretical considerations suggest that significant lopsidedness will begin to occur beyond 1.8 scale radii \citep{jog00}.  If one prefers a single value for the measure of lopsidedness, we recommend averaging $\langle A_1 \rangle_i$ and  $\langle A_1 \rangle_o$. 

\begin{figure*}[htbp]
\plotone{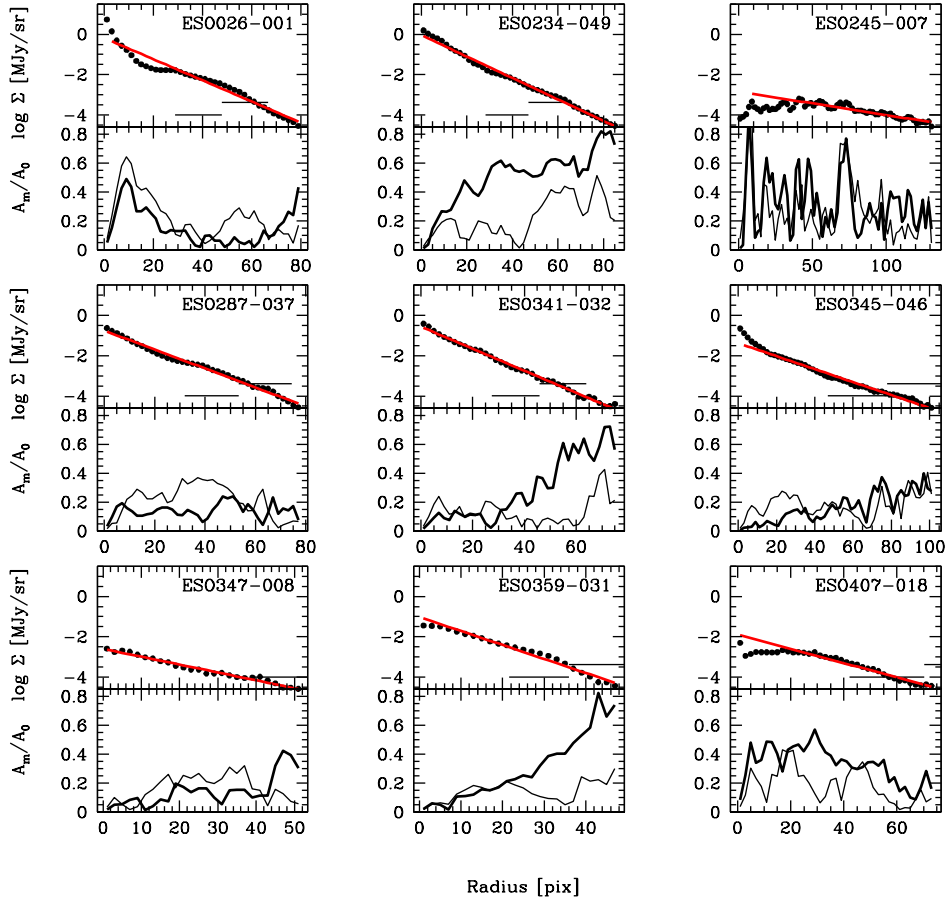}
\caption{Surface brightness, $\Sigma$, and $m=1$ and 2 Fourier component amplitude profiles for our sample of 167 galaxies (only first nine shown here, all are included in the electronic version of this Figure). Upper panels for each galaxy include the measured $A_0(r)$ and the results of our exponential fits (inclined solid red lines). The short horizontal lines demarcate the two radial ranges over which we calculate the average values of $A_1/A_0$ and $A_2/A_0$, $\langle A_1 \rangle_i$ and $\langle A_1 \rangle_o$ respectively. The radial behavior of  $A_1/A_0$ and $A_2/A_0$ are shown in the lower panels, in thick and thin lines respectively.}
\label{fig:a1}
\end{figure*}

To obtain a measure of a scaling radius, we develop a robust approach that is as independent of the underlying profile and multiple components as possible. This crude approach is necessary because our sample spans a large range of morphological types, including early types (E/S0). 
Rather than taking a detailed approach that attempts multi-component fits and profiles that have additional freedom, such as S\'ersic profiles \citep{sersic}, we do the following.
We fit an exponential to the mean radial surface brightness data, $A_0(r)$ using an iterative approach of $\chi^2$ minimization. On the first pass, we fit over all available radii. We then exclude data that are 0.2 dex above the fit and refit. This step helps remove sharply rising features, such as a central bulge or nucleus. We then repeat the fitting excluding only the data that are 0.4 dex above the fit. This step has the same aim, but uses the refined fit. Finally, we run two additional iterations that only exclude data that are 0.6 dex from the fit. These iterations remove outliers in either direction relative to the fit. Because not all of the surface brightness profiles are described well by exponentials \citep[see][for specific examples within S$^4$G]{jc}, the fits for certain galaxies are poorer than for others, although in general they are reasonably good and certainly indicative of how quickly the luminosity profile declines with radius (Figure \ref{fig:a1}). The goal of this approach is to provide a rough measure of scale (when we average over a range of radii  we are not sensitive to the exact measure of $R_S$) and one that is robust, simple, and can possibly be applied to simulations. This method is not intended to provide precise measures of the disk scale length. Those will be presented elsewhere for S$^4$G galaxies \citep{salo}. We retain the final $\chi^2$ values as a relative figure of merit of the fit and those values are included in Table \ref{tab:data} and used to color code galaxies in some of our Figures. Example fits for the first nine galaxies in our list are shown in Figure \ref{fig:a1} (all fits are shown in the electronically available version of this Figure). None of our subsequent results varies with the quality, $\chi^2$, of these fits. 

\begin{figure*}[htbp]
\plotone{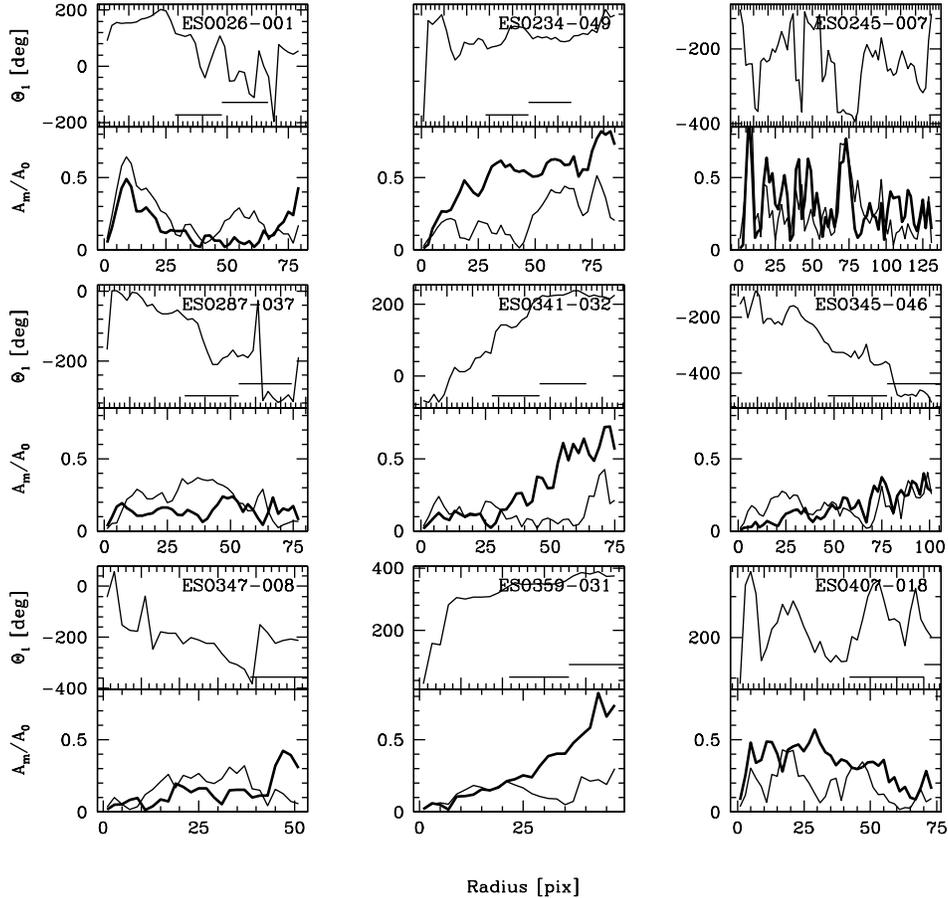}
\caption{Phase and $m=1$ and 2 Fourier component amplitude profiles for our sample of 167 galaxies (only first nine shown here, all are included in the electronic version of this Figure) . As a complement to Figure \ref{fig:a1} we show here the phase of the $m=1$ component as a function of radius for each galaxy in the upper panels. The lower panels are the same as those in Figure \ref{fig:a1}.}
\label{fig:phase}
\end{figure*}

\begin{figure}[htbp]
\plotone{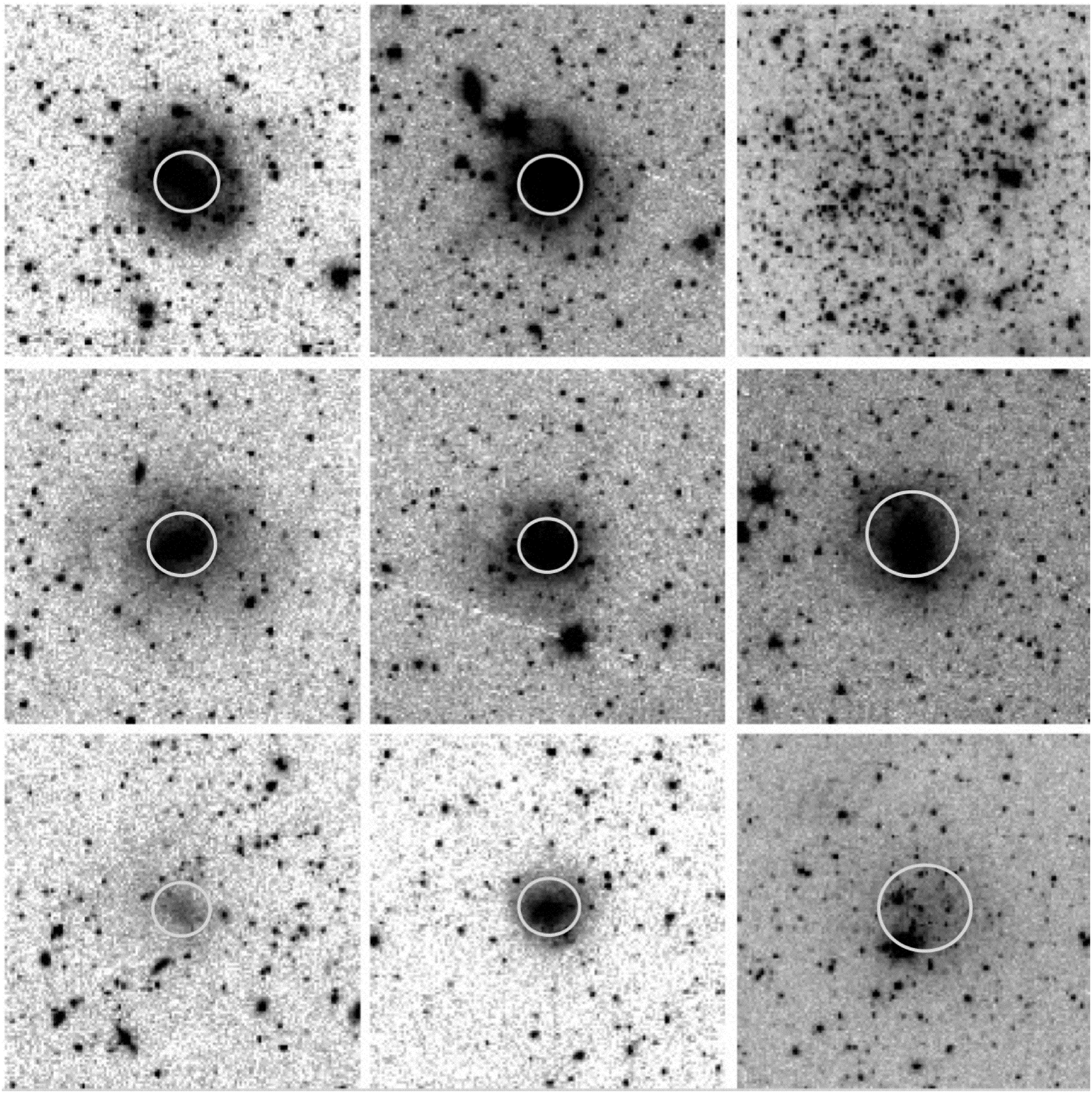}
\caption{As a complement to Figures \ref{fig:a1} and \ref{fig:phase}, we show images of the first nine galaxies and an indication of the inner edge of the inner radial range over which we measure the mean lopsidedness. The galaxy in the upper right corner (ESO245-007) is partially resolved and low surface brightness. Its surface brightness profile indicates that $R_S$ is too large and so it is not analyzed further. The image sections shown are all scaled similarly and correspond to only the inner 300 $\times$ 300 pixels (3.75 arcmin on a side) about the galaxies.}
\label{fig:lopimage}
\end{figure}

In intervals of two pixels in radius, 
we evaluate the Fourier components and phases up to $m=4$. The radial profiles of the amplitudes of the $m = 1$ and 2 components are shown in Figure \ref{fig:a1} and the corresponding phases for the $m=1$ components are shown in Figure \ref{fig:phase}. The average values over the defined radial intervals are listed in Table \ref{tab:data}.
Of the 186 galaxies we began with, only 167 have best fit radial profiles that indicate that the images we have completely include the radial range from 1.5 to 2.5 $R_s$. This slightly smaller sample is what we now discuss and include in the Tables and Figures.

\section{Results}

\subsection{The Radial Behavior of $A_1/A_0$}

\begin{figure*}[htbp]
\plotone{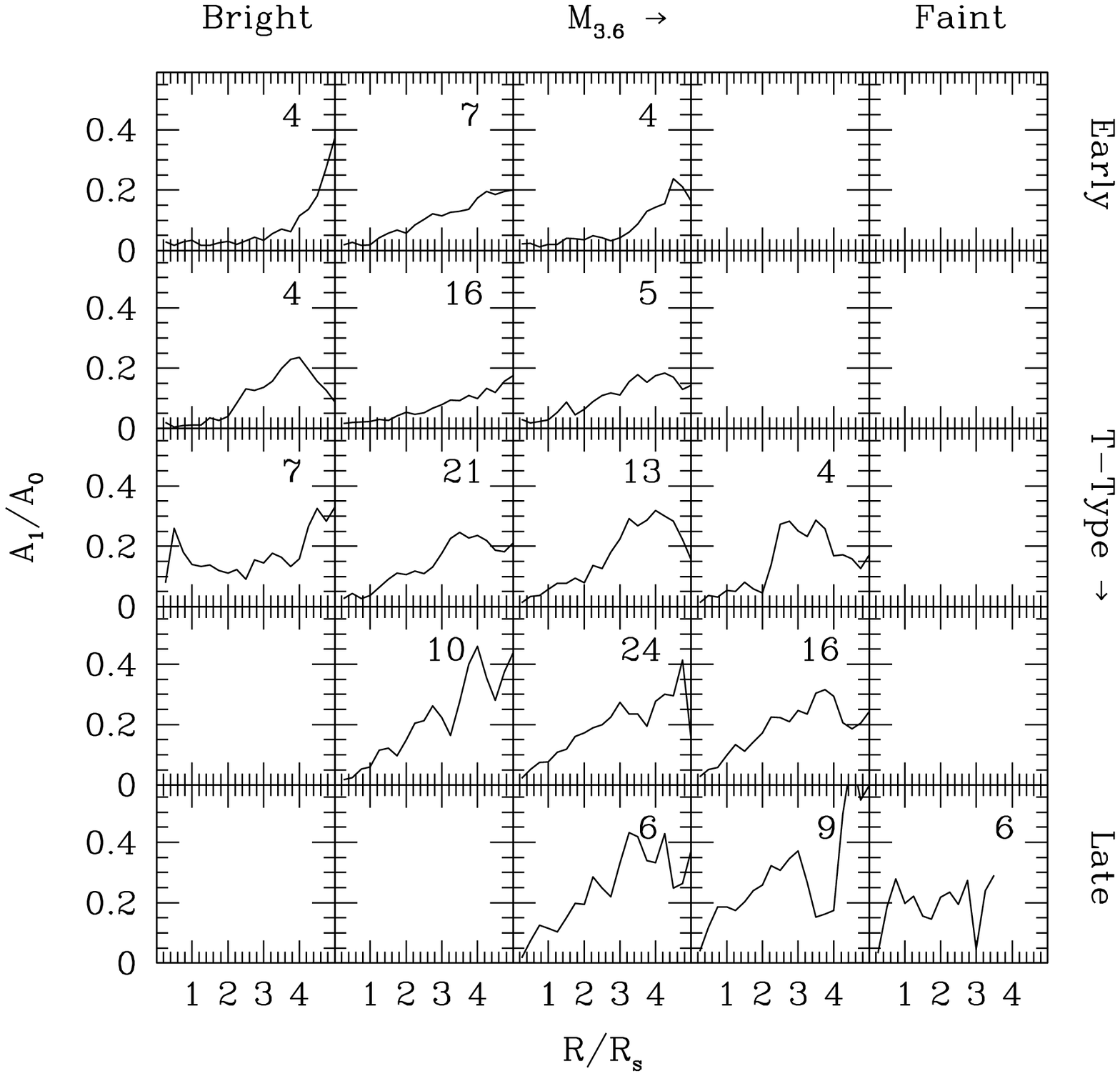}
\caption{Radial $A_1/A_0$ profiles. We present median profiles for bins in 3.6$\mu$m magnitude and T-type, when a bin contains 3 or more galaxies. Uncertainties are calculated from the distribution of values in a bin if it contains more than 5 measurements. Radius is given in units of scale lengths. The magnitude bins represent the range $-23$ to $-16$ evenly, while in T-type they represent $\le-1$, $(-1,-2]$, $(2,5]$, $(5,8]$, and $> 8$. The numbers in the panels indicate the number of galaxies contributing to that panel.}
\label{fig:profiles}
\end{figure*}

In Figure \ref{fig:profiles} we show the radial behavior of $A_1/A_0$ for galaxies as a function of their 3.6$\mu$m absolute magnitude \citep[the evaluation of the asymptotic total 3.6$\mu$m magnitudes, $M_{3.6}$, for the S$^4$G sample are described and presented by][]{jc} and morphology. Each panel represents $\ge 3$ galaxies, with a median calculated at a particular radius only if there are $\ge 5$ measurements within that radial bin. Two features stand out. First, among the early types (top two rows) there is very little lopsidedness inside of two scale lengths. For the brightest, presumably most massive, early types there is almost no sign of lopsidedness within 4 scale lengths. Strong $m=1$ distortions are therefore not to be expected in all galaxies at all radii. In particular, some property of these early type galaxies, at these radii, damps such distortions. Second, the $m=1$ amplitudes generally rise toward larger galactic radii independent of galaxy type or $M_{3.6}$ magnitude (and presumably mass). 
Even in those early type galaxies that show little or no $m=1$ component at smaller radii, it can be significant at larger radii. The possible exceptions to this broad statement are the faint very late type galaxies, where this mode can be large at all radii.  The trend toward greater amplitudes at larger radii demonstrates that centering issues are not the root cause of our $m=1$ distortions, as such errors would lead to larger $m=1$ amplitudes at small radii, and smaller amplitudes at large radii. The detection of rising distortions with radius has been noted before in other samples \citep{rudnick,conselice,reichard08}. Although one might suspect that this behavior simply reflects the longer dynamical times of matter at larger radii, it is also the signature of a model where lopsidedness arises from the asymmetry of the underlying dark matter potential, which has a more dominant role at larger radii \citep{jog99}. The trend in $m=1$ behavior with morphology suggests that dynamical times are not the only factor in determining the appearance of this mode.

\begin{figure}[htbp]
\plotone{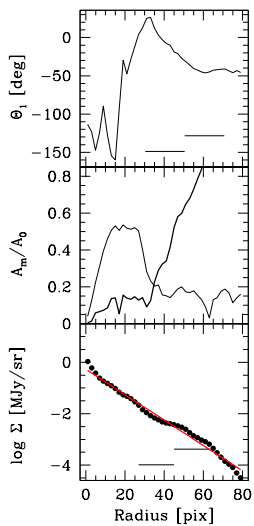}
\caption{Fourier parameters for NGC 3906. In the three panels we show the $m=1$ phase (upper), the relative $m=1$ amplitude (middle panel, thicker line) and $m=2$ amplitude (thinner line), and the surface brightness profile and fit. The $m=1$ component is clearly associated with lopsidedness in the outer radial interval, and is probably wound up in the inner radial interval. }
\label{fig:3906}
\end{figure} 

\begin{figure}[htbp]
\plotone{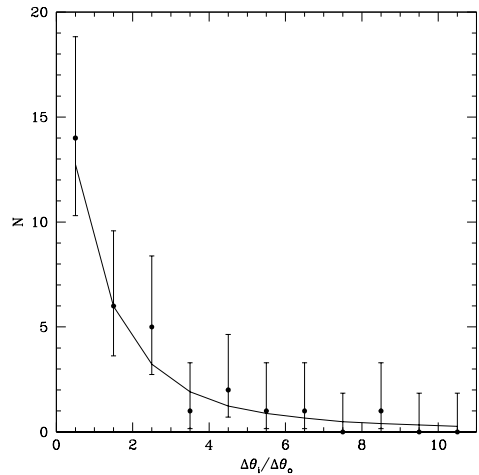}
\caption{The distribution of $m=1$ phase angle changes across the inner and outer radial intervals. The solid line represents the best fit model described in the text and demonstrates that observational uncertainties in combination with a small amount of winding can reproduce the observations.}
\label{fig:phase1}
\end{figure} 

\begin{figure}[htbp]
\plotone{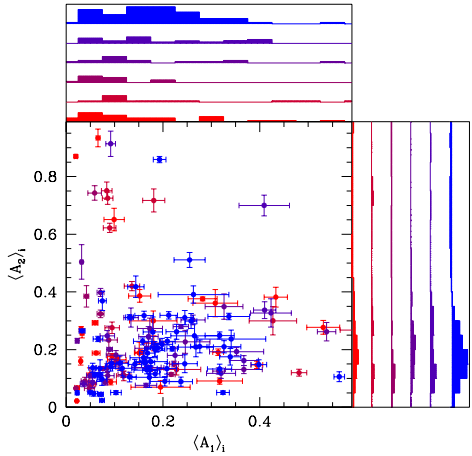}
\caption{ $\langle A_2 \rangle_i$  vs. $\langle A_1 \rangle_i$, representing the mean values for ($1.5 \le R/R_S \le 2.5$). Colors represent $\chi^2$ values in the radial profile fitting, with blue being the lowest $\chi^2$ or best quality,  and orange the highest or worst. The upper and right panels show the distribution of the galaxies along that axis for the different values of $\chi^2$.}
\label{fig:a1_inner}
\end{figure} 

\begin{figure}[htbp]
\plotone{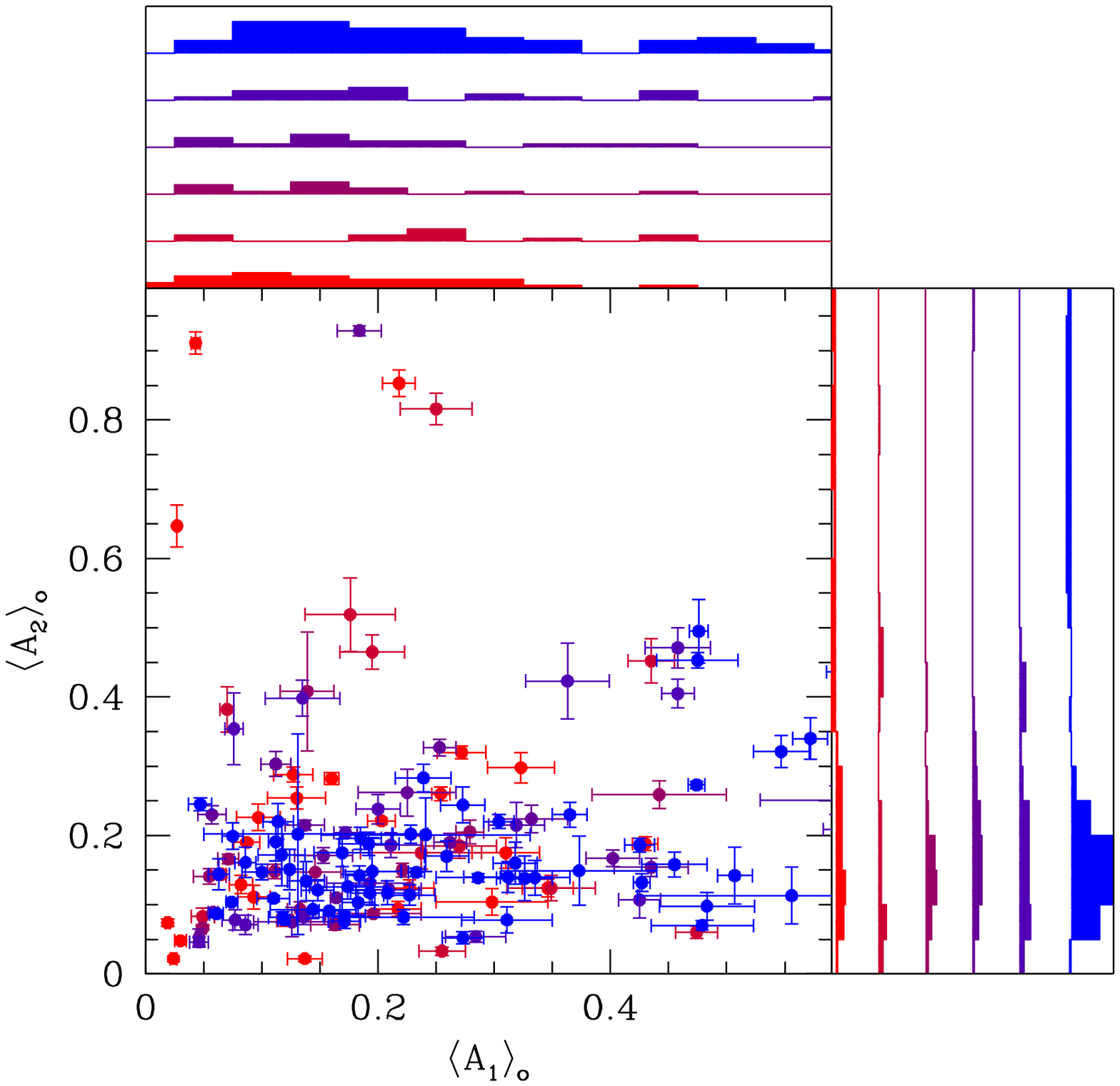}
\caption{$\langle A_2 \rangle_o$  vs. $\langle A_1 \rangle_o$, representing the mean values for ($2.5 \le R/R_S \le 3.5$). Colors represent $\chi^2$ values in the radial profile fitting, with blue being the lowest and orange the highest. The upper and right panels show the distribution of the galaxies along that axis for the different values of $\chi^2$.}
\label{fig:a1_outer}
\end{figure} 

\begin{figure}[htbp]
\plotone{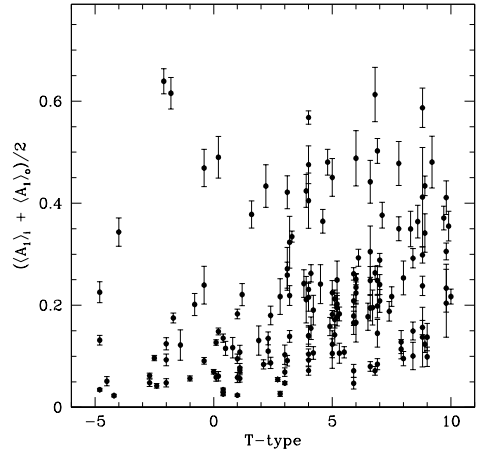}
\caption{Average of $\langle A_1 \rangle_i$  and $\langle A_1 \rangle_o$ vs. T-Type. One galaxy without an available type is excluded, as is one galaxy with a value of $\langle A_1 \rangle_i \sim 1.5$.}
\label{fig:type}
\end{figure}

\begin{figure}[htbp]
\plotone{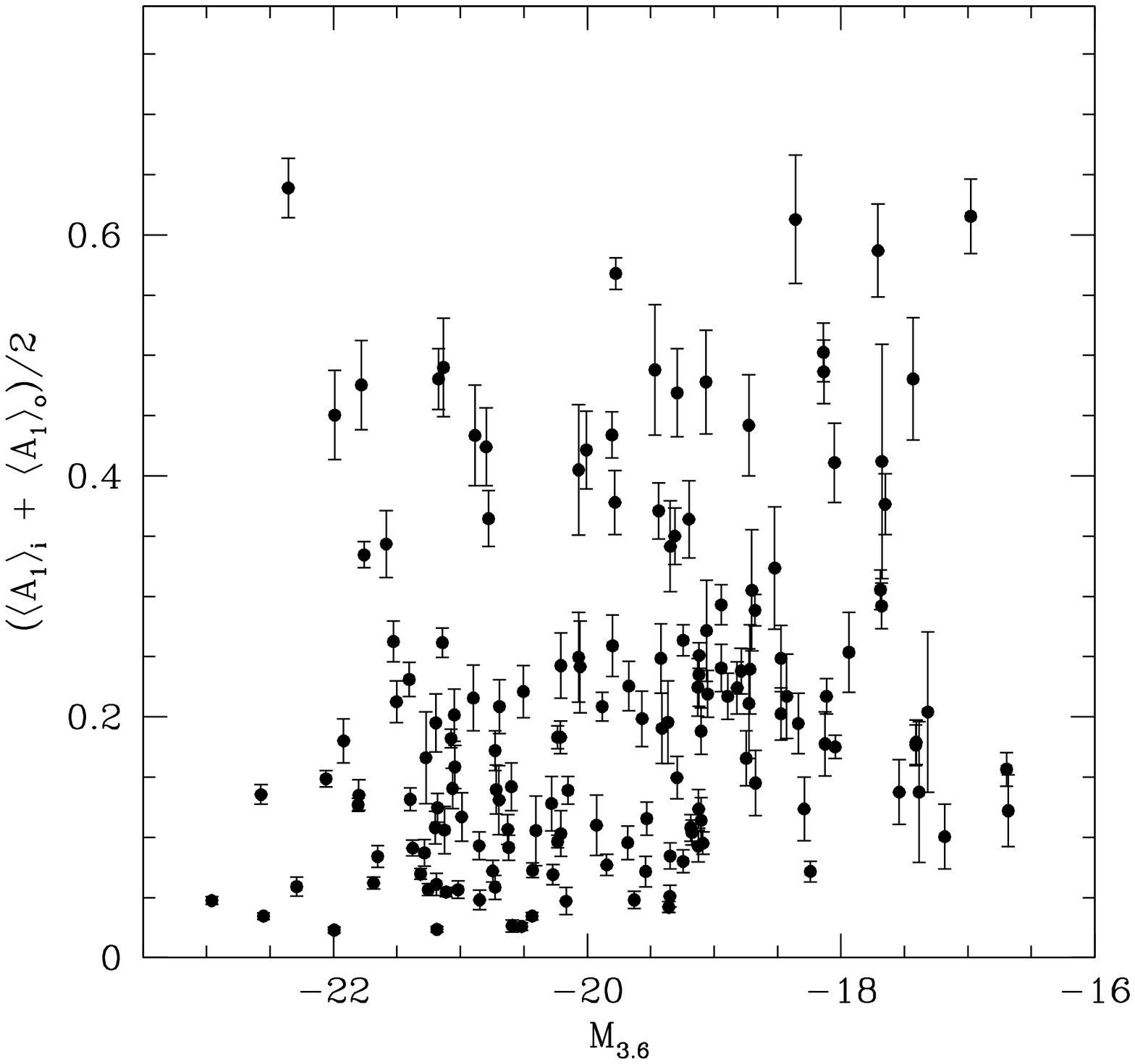}
\caption{Average of $\langle A_1 \rangle_i$  and $\langle A_1 \rangle_o$ vs. 3.6$\mu$m absolute magnitude. Two galaxies without $M_{3.6}$ measurements are excluded, as are three galaxies with $M_{3.6} > -15$ and one galaxy  with a value of $\langle A_1 \rangle_i \sim 1.5$}
\label{fig:mb}
\end{figure}

\subsection{The Morphological Nature of the $m=1$ Distortions}

In general, $m=1$ distortions are referred to as lopsidedness \citep[e.g.][]{rix,bournaud,reichard08}, even though this terminology is not quantitatively correct. An $m=1$ mode can wrap around in azimuth, resulting in a feature that resembles a one-armed spiral rather than lopsidedness. An example of this can be seen in the inner half of the $m=1$ component shown in Figure \ref{fig:decompose}.  As such, lopsidedness requires not only significantly large values of $A_1/A_0$, but also a level of constancy in the phase, $\theta_1$ \citep[see][for one example of a quantitative phase criteria]{li}. An analogous ambiguity exists in the interpretation of $m=2$ features, which although generally interpreted to imply the existence of a bar, can also arise from spiral arms.

We quantify the change in $\theta_1$ over our two chosen radial intervals by fitting a line to the $\theta_1(r)$ values within each interval. We define the $m=1$ distortion as lopsidedness if the phase profile slope, as evaluated by the fitted line, represents less than a 45$^\circ$ change in phase angle over the relevant radial interval.
To determine whether we are, in general, measuring lopsidedness, we consider the following.
Sixty and fifty-eight percent of the galaxies with large $\langle A_1 \rangle_i$,  $>$0.3, and well determined phase changes ($\sigma_{\Delta \theta_1} < 20^\circ$) satisfy $\Delta \theta_1 < 45^\circ$ within the radial ranges, $1.5 < R/R_s < 2.5$ and $2.5 < R/R_s < 3.5$, respectively. Requiring a phase change of less than 30$^\circ$ still results in nearly half of the galaxies being included (47 and 54\% respectively). We conclude that for at least half of the galaxies with large values of $\langle A_1 \rangle_i$, the correct geometrical description is indeed lopsidedness. Even the significant fraction of systems for which the $m=1$ distortion has a larger variation in $\theta_1$ may have started with true lopsided features that have been wound around by differential rotation.
For the example we show in Figure \ref{fig:decompose}, NGC 3906, we reproduce the relevant information in Figure \ref{fig:3906} from which it is clear that at large radii the $m=1$ phase is constant and confirms the visual impression of lopsidedness from Figure \ref{fig:decompose}. On the other hand, the nature of the $m=1$ distortion within the inner radial interval is complicated, with a significant phase swing, which can also be seen in the image of the galaxy. Other studies have also generally identified the $m=1$ mode at large radii as lopsidedness \citep{jog97, bournaud, angiras, eymerenb}.

\subsection{Winding Up $m=1$ Distortions}

One of the key questions regarding lopsidedness is the lifetime of such features. Because the lifetime, in combination with the rate at which these features are generated, helps determine the fraction of galaxies that have such a feature, it bears directly on the question of the physical origin of lopsidedness. In general, differential rotation will erase any structure quickly, and this has been appreciated in the context of lopsidedness since the first systematic study of the phenomenon \citep{baldwin}. The lifetime can be extended if there is either strong self-gravity \citep{saha} or an underlying asymmetric potential, and so a theoretical estimate of the lifetime of lopsidedness is difficult to calculate from first principles. 

We use the measured change in phase as a function of radius to constrain the amount of winding.
We begin with the $\theta_1$ vs. radius linear fits in each of our two radial ranges that we calculated previously. In the case of a constant circular velocity, the ratio of the phase changes across the two radial ranges due to kinematic winding can be calculated (2.3, with the higher phase change across the inner annulus). The magnitude of the phase change across two radii depends on the circular velocity and time since the feature was generated (assuming that no phase change was present across the original feature), but this {\sl ratio} is independent of the circular velocity and the physical units of radius. The distribution of this ratio for our galaxies is shown in Figure \ref{fig:phase1}. It does not peak at 2.3, indicating either that the phase change behavior across radius is not due to purely kinematical effects or that measurement uncertainties have worked to erase the peak. The smaller the actual phase changes (either because the circular velocity is small, the time since the feature was created is short, or something is counteracting the winding) the easier it is to alter the peak for a given 
measurement uncertainty (while the ratio is fixed, the effect of uncertainties on this ratio does vary). For our data, the formal slope uncertainties tend to be $\sim$ 15$^\circ$.

Using a Monte Carlo approach to simulate the effects of uncertainties, we test for the range of acceptable phase changes in the outer annulus (the inner one then has a change that is 2.3 times larger under the assumption of differential rotation and a flat rotation curve). For an error of 15$^\circ$, we find that the 90\% confidence interval on the allowed phase change in the outer radial bin is between 18.6$^\circ$ and 37.4$^\circ$. Therefore,  the data allow for some ``winding" of the $m=1$ feature consistent with differential rotation. For a typical rotation velocity of 150 km sec$^{-1}$ and a scale length of 2 kpc, a phase difference of 37$^\circ$ would develop in 
slightly over $8\times10^7$ yrs. Even in later type galaxies, where the rotation curves are more often characterized as rising than flat, the rotation velocities drop well short of a solid body rise beyond the innermost radii (solid body rotation results in no winding).  An increase by a factor of a few in winding times does not resolve the kinematic winding problem. Therefore, the distortions are either produced often or  they last much longer than this estimate due to other physical effects. If the distortions are produced often then they would need to be produced about  every $10^8$ yrs in each galaxy, so as to be seen in a large fraction of all galaxies. This timescale is too short for external events because we do not expect an accretion event or flyby of a satellite galaxy every $10^8$ years. So either the self-gravity is important, or there is a continuous forcing, such as from a bar of an offset halo.
The conclusion that kinematic winding is not dominant has been reached before \citep{baldwin,ideta}, although here we place a more quantitative limit on the expected lifetime if differential rotation is the sole operating factor and detect some evidence for slight winding due to differential rotation.

\subsection{Statistical Properties of $\langle A_1 \rangle$ and $\langle A_2 \rangle$}

In Figures \ref{fig:a1_inner} and \ref{fig:a1_outer}, we show the relationship between $\langle A_1 \rangle$ and $\langle A_2 \rangle$ for our two radial ranges. For the inner radial range, the two quantities often, but not exclusively, rise in concert, confirming again that values of $\langle A_1 \rangle_i$ are not the result of a systematic error (such as miscentering). We color code the galaxies by the quality of the surface brightness profile fit used to derive $R_S$ to highlight any potential dependence arising from our exponential fitting of the surface brightness profiles. We find no evidence that any of the following results depend on the quality of this fitting. 

Although many of the galaxies lie nearly on the 1:1 line between the two axes in Figure \ref{fig:a1_inner}, there are two populations that deviate significantly from this relation. First, a subset of galaxies with large ($>$0.4) values of  $\langle A_2 \rangle_i$ is present.  From visual inspection,  we conclude that in general these are systems with strong bars, which cause the high values of  $\langle A_2 \rangle_i$. Unlike the galaxy in Figure \ref{fig:decompose}, strong bars are not always associated with strong $m=1$ features because most bars are centered and roughly symmetric. 
We explore this issue further in \S3.5. Second, there is a tail of systems to large ($>$ 0.3) values of  $\langle A_1 \rangle_i$. Again, we visually examined these and find that nearly one half of these systems appear to be actively interacting. A few more have either a poorly defined center or a bright knot near the center that could have been mistaken for the center. We conclude that the majority of these systems are not the class of interest. Overall, the results suggest that within the inner radial range, $m=1$ asymmetries can be closely related to $m=2$ asymmetries. This correlation does not arise from a single phenomenon. For example, one subpopulation that satisfies this trend is that of the Magellanic Irregulars that, like the Large Magellanic Cloud \citep[see,][]{zar04}, tend to have an off-center bar \citep{deV} and about half the time enhanced star formation at one end of the bar \citep{ee}, which can account for both high $m=1$ and 2 amplitudes \citep{colin}, but these represent a small fraction of the total galaxy sample.

At larger radii, Figure \ref{fig:a1_outer}, the behavior is different, with less appreciable correlation between  $\langle A_1 \rangle_o$ and  $\langle A_2 \rangle_o$.  The data seem to suggest that the $m=1$ asymmetries persist to larger radii, and even grow in relative amplitude. This phenomenon has several possible origins. For example, $m=2$ modes may be highlighted by the star formation that occurs in arms. Although the sensitivity to star formation is diminished at these wavelengths relative to optical bands, significant flux ($\sim$ 30\%) can come from young, massive stars in these passbands \citep{meidt,eskew}. Beyond the star formation threshold \citep{kennicutt}, the amplitude of these asymmetries, even if not decreasing in relative mass, would be diminished. Alternatively, the $m=2$ mode may not propagate beyond some critical resonance, such as the outer Lindblad resonance, although weak spiral structure is visible in the distribution of outer disk HII regions and star clusters \citep{ferguson,thilker,christlein,shf}.
Interestingly, a visual study of outer disks in this sample found no correlation in lopsidedness at large radii and interactions or the presence of nearby companions \citep{seppo}, potentially providing another important clue to the origin of these features.

\subsection{Connections to Morphology and Luminosity}

To increase the signal-to-noise of our lopsidedness measure, we now average the inner and outer radial interval measures.
Plotting  the value of ($\langle A_1 \rangle_i + \langle A_1 \rangle_o)/2$ relative to galaxy morphological type in Figures \ref{fig:type}, we find a well defined upper envelope 
that increases steadily toward later types. The correlation with type (Spearman rank correlation coefficient of 0.397, corresponding to a probability of arising at random of $1.5\times 10^{-7}$) is more the result of this envelope than a well-defined correlation because even for the latest types some galaxies exist with low values of  ($\langle A_1 \rangle_i + \langle A_1 \rangle_o)/2$. The lack of a strict correlation suggests that the physical origin of this effect may be related to a characteristic that correlates with morphological type, but varies within any given T-type. For example, \cite{reichard08} found a strong correlation between lopsidedness and surface density. Because the earlier types are generally of higher surface density, such a correlation would give rise to an effect that is noticeable among galaxy morphologies as well. Of course, there are other  galaxy properties that correlate with morphology, such as gas richness, environment, and bulge mass for example, that could also play a role in the presence or  absence of lopsidedness. 
 
Morphology, or some property that correlates with morphology, is a somewhat stronger indicator of lopsidedness than stellar luminosity (or mass). 
In Figure \ref{fig:mb}, we plot  ($\langle A_1 \rangle_i + \langle A_1 \rangle_o)/2$ vs. $M_{3.6}$. Although the visual impression is significantly less striking than in Figure \ref{fig:type}, a significant correlation does exist (correlation coefficient of 0.312 and a probability of arising at random of $6.3\times 10^{-5}$) but the quantitative results suggest that it is somewhat weaker than that with type. It is perilous to compare correlation strengths between quantities with different uncertainty distributions. 

\begin{figure}[htbp]
\plotone{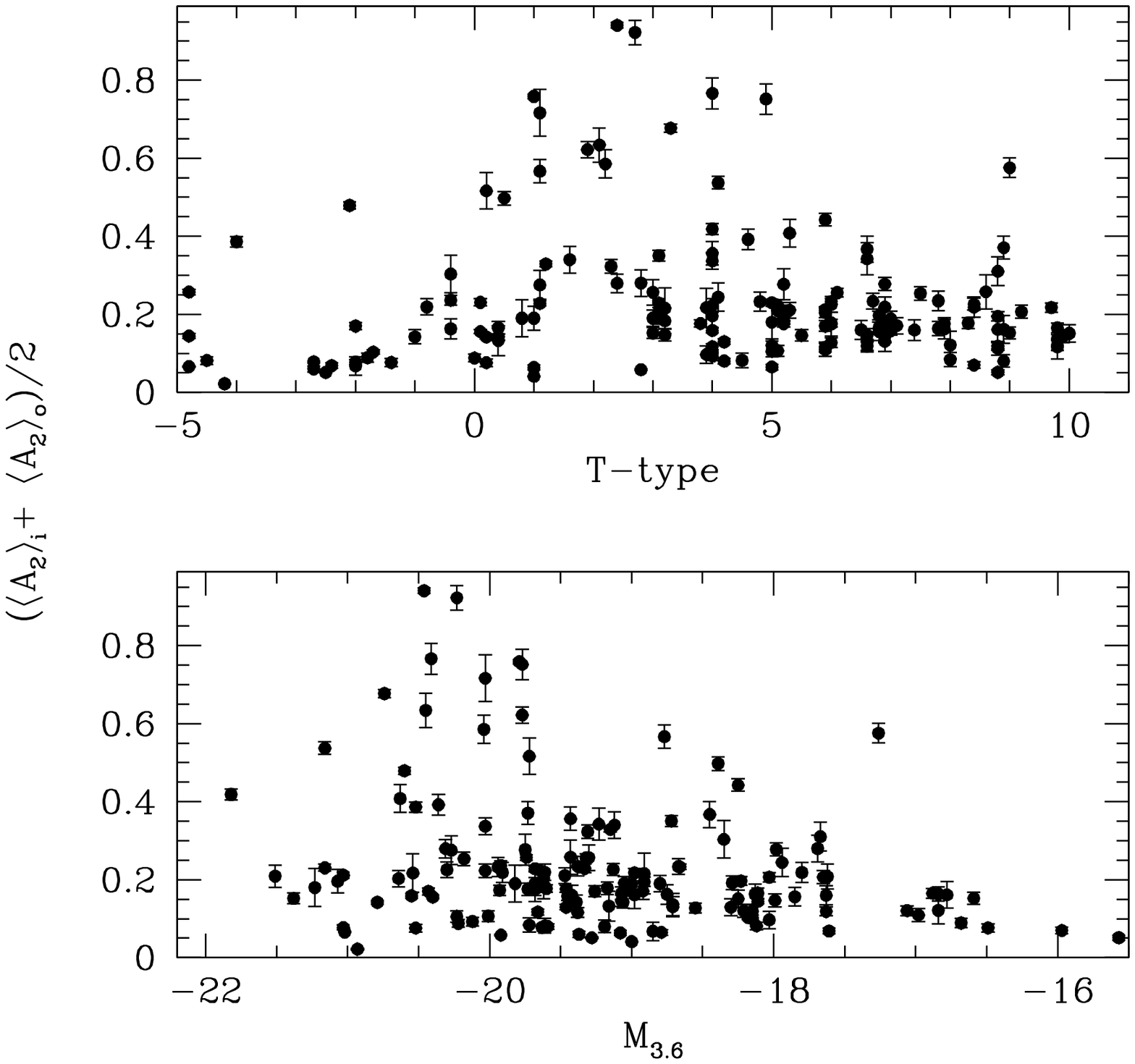}
\caption{Average of $\langle A_2 \rangle_i$  and $\langle A_2 \rangle_o$ vs. T-Type (upper panel) and 3.6$\mu$m absolute magnitude (lower panel). }
\label{fig:a2}
\end{figure}

Interestingly, the behavior of $\langle A_2 \rangle_i$ is quite different than that of $\langle A_1 \rangle_i$, even though we saw before that the two are often related. First, in the upper panel of Figure \ref{fig:a2} we show that while  $\langle A_2 \rangle_i$ does not correlate with type, the strong values of  $\langle A_2 \rangle_i$ are limited to a narrow range of T-Types (1 -- 5). Otherwise, the strength of the $m=2$ modes is fairly T-type independent (although recall that this mode measures both bars and spiral arms). Second, in the lower panel of Figure \ref{fig:a2}, we show that high $m=2$ amplitudes are mostly contained near $M_{3.6} \sim - 20.5$. Again, other than those large  $\langle A_2 \rangle$ values, the distribution appears to be independent of $M_{3.6}$. So, unlike $m=2$ distortions, which are generally considered to be common but are actually confined to specific masses and morphologies, $m=1$ distortions, which are generally considered to be somewhat special, can occur in any galaxy despite the preference for later types. 

The results above are coarse in that they combine a set of different asymmetries, bars and arms, into one category. We now examine how lopsidedness behaves with respect to arm and bar classifications. We classify the arm and bar types as done for a subset of S$^4$G galaxies by \cite{elmegreen}, specifically categorizing galaxies with spirals arms as flocculent, multiple arm, or grand design, and all disks as either having a weak, a strong, or no bar. The results are included in Table \ref{tab:data} and the statistics are presented in Tables \ref{tab:arms} and \ref{tab:bars}.

\begin{deluxetable*}{lrrrrr}
\tablecaption{Arm Class Dependence}
\tablewidth{0pt}
\tablehead{
\colhead{Arm Class}&
\colhead{$\langle A_1 \rangle_i$}&
\colhead{$\langle A_2 \rangle_i$}&
\colhead{$\langle A_1 \rangle_o$}&
\colhead{$\langle A_2 \rangle_o$}&
\colhead{Number} \\
}
\startdata
Flocculent&0.28$\pm$0.03&0.21$\pm$0.02&0.35$\pm$0.03&0.19$\pm$0.03&50\\
Multiple Arm&0.16$\pm$0.01&0.23$\pm$0.02&0.22$\pm$0.02&0.20$\pm$0.02&72\\
Grand Design&0.12$\pm$0.02&0.48$\pm$0.07&0.21$\pm$0.04&0.37$\pm$0.06&26\\
\enddata
\label{tab:arms}
\end{deluxetable*}

\begin{deluxetable*}{lrrrrr}
\tablecaption{Bar Class Dependence}
\tablewidth{0pt}
\tablehead{
\colhead{Bar Class}&
\colhead{$\langle A_1 \rangle_i$}&
\colhead{$\langle A_2 \rangle_i$}&
\colhead{$\langle A_1 \rangle_o$}&
\colhead{$\langle A_2 \rangle_o$}&
\colhead{Number} \\
}
\startdata
None&0.20$\pm$0.02&0.20$\pm$0.02&0.27$\pm$0.02&0.20$\pm$0.02&95\\
Weak&0.22$\pm$0.02&0.22$\pm$0.02&0.25$\pm$0.02&0.18$\pm$0.02&29\\
Strong&0.15$\pm$0.02&0.40$\pm$0.04&0.24$\pm$0.03&0.30$\pm$0.04&43\\
\enddata
\label{tab:bars}
\end{deluxetable*}

Beginning with the arm classification, there is a strong dependence between the arm class and both $\langle A_1 \rangle_i$ and $\langle A_2 \rangle_o$. Flocculent spirals have the strongest $m=1$ distortions, while the grand design spirals have the weakest. In a scenario where grand design spirals, such as that in M 51, are initiated by close a fly-by, this argues against such an event being critical to the creation of most $m=1$ distortions. This interpretation would agree with the finding that lopsided galaxies are generally not obviously interacting systems or accompanied by nearby companions \citep{jog,seppo}.  Likewise, these results argue that the same physical characteristic of a galaxy that promotes an $m=1$ mode is related to the flocculent nature of the spiral arms. 

In contrast, the presence of a bar seems to have only a modest effect on the $m=1$ properties, with a statistically significant difference only for strong bars and $m=1$ in the inner radial range. This result appears at odds with the correlation seen in Figure \ref{fig:a1_inner} where, for the inner radial range, $\langle A_1 \rangle$ and $\langle A_2 \rangle$ correlate. Interpreting that correlation, however, is difficult. First,  larger $\langle A_2 \rangle$ can be a signature of a bar, but it can also reflect the presence of spiral arms. Second, the correlation seen between morphology and lopsidedness is also present to some degree between morphology and  $\langle A_2 \rangle$. The earliest type galaxies, ellipticals and lenticulars, have little or no  $\langle A_1 \rangle$ (Figure \ref{fig:profiles}) and by definition no spiral arms. We therefore expect some correlation between  $\langle A_1 \rangle$ and  $\langle A_2 \rangle$ based on the range of morphologies in the sample. The least ambiguous data to examine for a connection between lopsidedness and bars is that of our bar classifications.
We conclude that the creation of a bar, whether as a result of an interaction or a disk instability, does not generally result in $m=1$ distortions, particularly such distortions at large radii. Conversely, the creation of an $m=1$ distortion does not generally result in a corresponding bar.

\begin{figure}[htbp]
\plotone{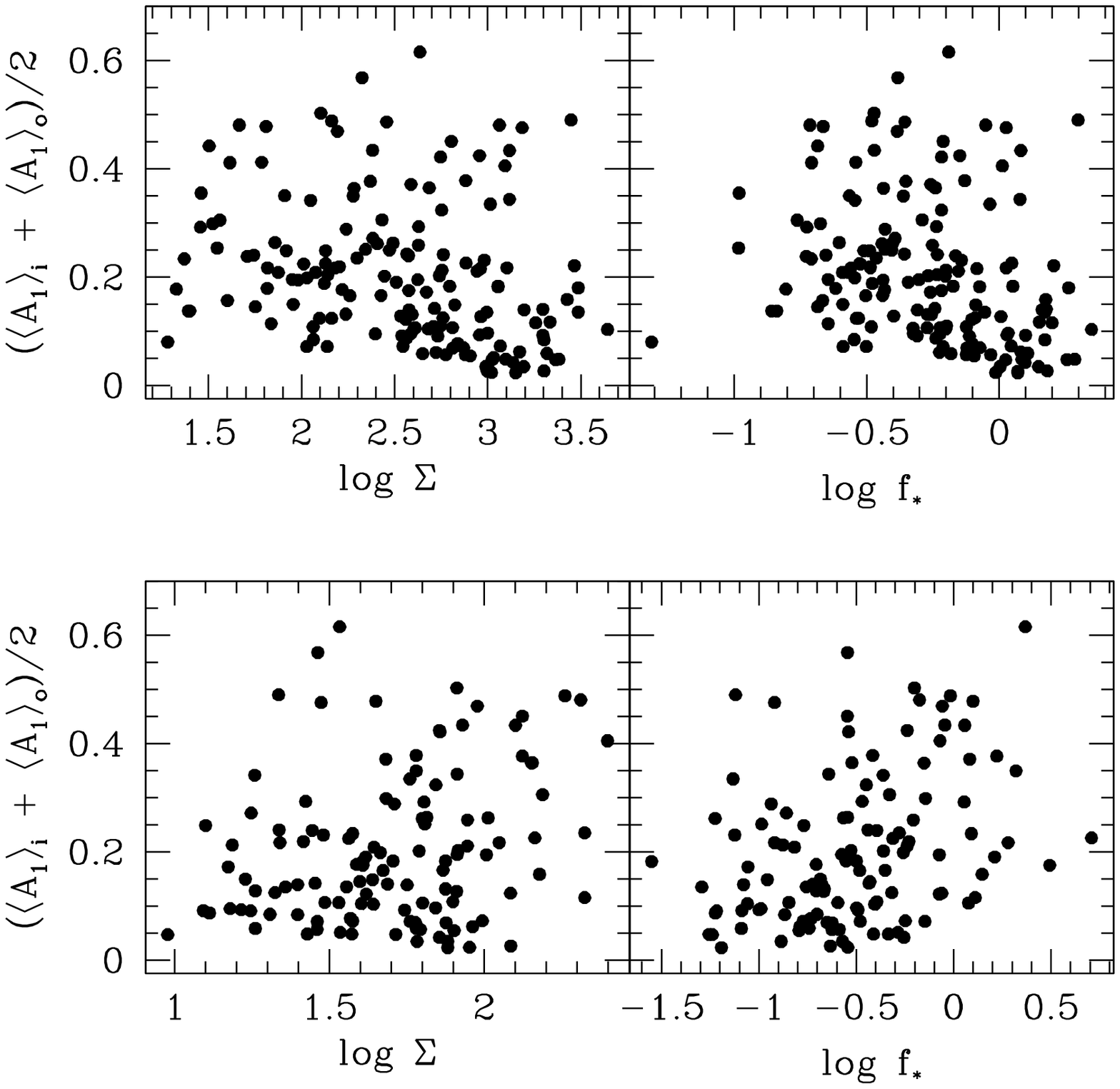}
\caption{Relationship between $m=1$ amplitudes measured either interior to $R_S$ (upper panels) or between 1.5 and 3.5 $R_S$ (lower panels) and stellar surface mass density, $\Sigma$, (left panels) and stellar mass fraction, $f_*$, (right panels). The correlations are stronger with 
$f_*$ and are inverted from one radial range to the other. }
\label{fig:sb}
\end{figure}

\subsection{Determining Stellar Masses, Surface Densities, Total Enclosed Masses, and Stellar Fractions}

Following the suggestion of \cite{rudnick} and \cite{reichard08}, we now explore the role of surface density on lopsidedness. In particular, we use our 3.6$\mu$m photometry to measure stellar surface density and, in combination with the IR Tully-Fisher relation that provides an estimate of the rotational velocity and therefore halo mass, the stellar mass fraction. 
First, we calculate the stellar mass using the transformation from $3.6\mu$m flux to stellar mass presented by \cite{eskew}. Specifically, the stellar mass, in solar units, can be expressed as $10^{5.9}F_{3.6}(D/0.05)^2$, where $F_{3.6}$ is the flux at 3.6$\mu$m in units of Jy and D is the distance to the galaxy in Mpc. Stellar surface densities are then calculated by dividing the stellar mass by the corresponding area over which the flux was measured. We measure the flux by integrating the values of $A_0(r)$ over the desired radial range. We obtain the rotational velocity, $v_c$, by inverting the Tully-Fisher relation at 3.6$\mu$m presented by \cite{sorce}. The enclosed total mass at any radius is then estimated using $v_c^2r/G$. The stellar mass fraction, $f_*$, is then defined to be the stellar mass over the radial range of interest divided by the total mass enclosed within the same radial interval. Despite the various simplifying assumptions involved in the calculation of both the stellar and total masses, the stellar fractions obtained are within a reasonable range, $0.1 < f_* < 1$, for the vast majority of the galaxies. A lower limit on the uncertainties in $f_*$ can be gauged by the spread of values above the physical limit, $f_*=1$.

In Figure \ref{fig:sb} we show the dependence of ($\langle A_1 \rangle_i + \langle A_1 \rangle_o)/2$ on both stellar surface density, $\Sigma$ in units of $M_\odot / pc^2$ and the stellar mass fraction, $f_*$, for radii inside $R_S$ and radii over which ($\langle A_1 \rangle_i + \langle A_1 \rangle_o)/2$ is measured, $1.5 R_S < r < 3.5 R_S$, for all galaxies with both $\langle A_1 \rangle_i$ and $\langle A_1 \rangle_o < 0.8$. While all panels show some evidence for correlations, an interesting trend is that the strength of $m=1$ distortions anticorrelates with $\Sigma$ and $f_*$ at small radii and correlates with the same quantities as measured at larger radii. Specifically, the inverse relationship with $\Sigma$ interior to $R_S$ is fairly strong (correlation coefficient, $R$, $-$0.354, probability of randomly arising, $P$, 5.2$\times10^{-6}$) and inverted and much weaker  over the radial range in which we measure lopsidedness ($R = $0.211, $P = 0.021$).  The trends with $f_*$ are stronger, even though to calculate $f_*$ we introduce a second set of assumptions and uncertainties that one might have expected to weaken the apparent correlations. Instead, the anticorrelation seen at small radius is very strong ($R = -0.387, P=5.2\times10^{-7}$) as is the positive correlation at larger radii ($R=0.353, P = 7.7 \times 10^{-5}$). 

At this point, when we are about to begin to interpret these correlations, it is critical to articulate the nature of the interdependencies between parameters. Specifically, both $\Sigma$ and $f_*$ correlate with morphology and $f_*$, as calculated, depends directly on $\Sigma$. Specifically, early T-Types tend to be more centrally concentrated and have higher $f_*$ in the inner regions and lower $f_*$ in the outer regions. Therefore, the measured correlations could be the result of another physical characteristic of galaxies that correlates with morphology. For example, if lopsidedness is more likely to occur in gas rich galaxies then we would expect a correlation with morphology that would give rise to a correlation with $\Sigma$ and  $f_*$. The often stated admonition ``correlations do not imply causality" is valid.

In an effort to identify whether the connection to $f_*$ is more fundamental than with T-type or $\Sigma$ we search for correlations between the residuals about the various mean trends with lopsidedness. Unfortunately, we find no compelling evidence for the primacy of one of these parameters over another in relation to the central concentration of stellar mass. There is a suggestion, as noted above, that $f_*$ within $R_S$ correlates slightly more strongly with lopsidedness than $\Sigma$ within this same radius, and this result is perhaps more significant given the crudeness of the $f_*$ calculation. Calculating $f_*$ using observed rotation curves might help clarify this distinction. On the other hand, we find that the correlation between $f_*$ measured over the radial range in which we measure lopsidedness ($1.5 R_S < r < 3.5 R_S$) and lopsidedness is noticeably superior to that with either $\Sigma$ or morphology. First, the correlation between lopsidedness and $\Sigma$ measured over this radial range is not significant. Second, when we limit the morphological types of galaxies considered to normal disks ($3 \le T < 8$), we retain a comparably strong correlation coefficient (0.331) that, despite the smaller sample size, remains significant ($P = 0.007$).

On the basis of the correlations between lopsidedness and $f_*$, we hypothesize that the strength of $m=1$ distortions for most galaxies depends on the interplay of the central relative concentration of stars to dark matter, with higher concentrations helping to dampen such distortions, and the outer relative concentration of stars to dark matter, with higher concentrations helping generate or sustain such distortions. This scenario, along with other results noted above, implies that lopsidedness for most galaxies is an internal phenomenon, perhaps depending on the asymmetry and centering of the dark matter potential or on a gravitational instability that is partially, but incompletely, stabilized by the dark matter halo or a dynamically hot stellar component. In the former, it is reasonable to expect that a higher concentration of stars to dark matter will help center the stellar and dark matter distributions, eliminating offsets that could have given rise to lopsidedness \citep{levine, noordermeer} and that these same stars will help circularize any asymmetry in the underlying potential \citep{jog99,bailin}, which could also have lead to lopsidedness \citep{weinberg,jog97}. If the stellar component is less dominant at small radii, these balancing effects would be lessened.
Suggestions of stabilization against lopsidedness by central concentrations, such as a bulge, have been made previously \citep{jog} and such concentrations may also diminish the expression of lopsidedness \citep{heller}. At larger radii,  because self-gravity reduces the effect of kinematic winding \citep{saha}, it is necessary that the stars represent a significant portion of the overall potential over the radii where the $m=1$ distortions are measured if they are to assist in the survival of any features for long periods. Analogous conclusions exist regarding the nature of spiral arms and the behavior of rotation curves \citep{ee90}. In the latter, it is primarily the large outer disk values of $f_*$ that are responsible for generating an m=1 mode of the flavor originally proposed by \cite{ostriker} to account for stellar bars.

This speculation is not intended to entirely exclude accretion, mergers, and or interactions as a potential source of $m=1$ distortions \citep{zaritsky, kornreich, mapelli}. In particular, in the upper panel of Figure \ref{fig:sb} one can perhaps identify two populations. First, there is the dominant one that appears to have increasing 
($\langle A_1 \rangle_i + \langle A_1 \rangle_o)/2$ with decreasing $f_*$. Second, there is a population with ($\langle A_1 \rangle_i + \langle A_1 \rangle_o)/2 > 0.3$ that is found at all values of $f_*$. These larger distortions may indeed be caused by interactions, while the lower level ones arise from internal halo asymmetries. Interestingly, the two populations are not distinct in the lower right panel of Figure \ref{fig:sb}, the one that probes the connection between lopsidedness and the mass fraction at the radii over which lopsidedness is measured, which may be due to the requirement that independent of the origin a substantial $f_*$ is needed to maintain a distortion. Independent evidence in the form of larger values of $m=1$ modes for galaxies in the Eridanus group \citep{angiras} also suggest that interactions do play some role, and other lines of evidence also support a tidal origin \citep{eymerenb}, even if only to make the dark matter potential asymmetric in the first place. Neither do we argue against asymmetric gas accretion as a cause of lopsidedness in the H \small{I} distribution \citep{kornreich01,bournaud, keres}, which would primarily affect the H {\small I} distribution at even larger radii \citep[such as explored by][]{eymerenb}.

If we accept the suggestion that moderate levels of lopsidedness reflect asymmetries in the underlying halo, we can connect ($\langle A_1 \rangle_i + \langle A_1 \rangle_o)/2$ with halo ellipticities, $\epsilon_h$. \cite{jog} calculate how the distortions in the isophotes are magnified versions, by a factor of $\sim 4$, of those in the potential.  Therefore, what we have termed moderate lopsidedness corresponds to $\sim$ 5\% distortions in the dark matter halo. It seems difficult to imagine how such small deviations from symmetry could be avoided in a hierarchically constructed halo. As such, it seems that any other phenomenon leading to lopsidedness, such as internal disk instabilities, would work as a supplement to that generated by halo lopsidedness.

\subsection{Kinematic Signatures}

In detail, the various origin scenarios can lead to different kinematic signatures. Unfortunately,  lopsidedness is most easily seen in face-on galaxies, such as those presented here, which are the least favorable for detecting kinematic deviations from a symmetric rotation curve. For example, \cite{rix} estimated the deviations from circular velocity expected for a potential that is mildly lopsided, assuming closed streamlines
$$\langle v_r \rangle = 7.4 {\rm \ km \ s}^{-1} \left({{v_c\over 200 {\rm \ km \ s}^{-1}}}\right) \left({\langle A_1 \rangle_i \over 0.11}\right) \left({2.5 R_S \over R}\right) \eqno{(2)}$$ with
comparable excursions in $v_\theta$. 
Effects at $\sim$ 10 km s$^{-1}$ are measurable in edge-on galaxies particularly at larger radii where extinction is less of a factor and the lopsidedness is larger. More detailed calculations suggest that asymmetries of 10\% or 20 to 30 km s$^{-1}$ should be expected \citep{jog97, jog02}. Such kinematic lopsidedness has been observed to be common, at least in the gas \citep{richter,swaters,eymeren}.

\begin{figure}[htbp]
\plotone{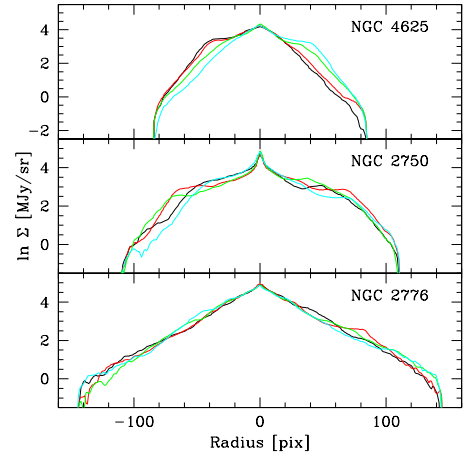}
\caption{Calculated edge-on surface brightness profiles for three of our face-on galaxies. Different lines in each panel represent a different viewing angle within the disk plane. The viewing angles are evenly spaced to cover all non-redundant angles. The three galaxies chosen are all disk galaxies with well-defined spiral patterns and weak to non-detected bulges and sample a range of $\langle A_1 \rangle_i$.}
\label{fig:edge-on}
\end{figure}

The surface brightness profiles of lopsided galaxies even when seen edge-on are detectably asymmetric if the viewing angle is favorable. To demonstrate this claim we have taken some of our face-on galaxies and effectively inclined them to our line of sight by summing the luminosity through the disk along lines-of-sight located within the disk plane. We chose three disk galaxies with insignificant bulges, for simplicity, and varying degrees of lopsidedness. This assumes no additional extinction in the disk plane, but even if this assumption is questionable, the extinction is unlikely to be such that it preferentially  removes asymmetries. In Figure \ref{fig:edge-on} we show three galaxy profiles, where we take 4 different viewing angles that are all in the disk plane, spaced equally in azimuth. Although asymmetries are visible in all cases, NGC 4625, with $\langle A_1 \rangle_i = 0.245 \pm 0.024$, has strong asymmetries where the surface brightness profile differs by about 2 magnitudes from one side to another when viewed from certain orientations. NGC 2750, with $\langle A_1 \rangle_i = 0.085\pm0.12$ shows less noticeable asymmetries, while the asymmetries for NGC 2776, with its larger $\langle A_1 \rangle_i = 0.194\pm0.018$ shows the least striking asymmetries. The last two demonstrate how sharp breaks in the profile can highlight asymmetries that are less pronounced in a purely exponential profile. Quantitatively this level of asymmetry should be recovered in the exponential profile, but asymmetries in the galaxies with sharp breaks are easier to select visually.

\begin{figure}[htbp]
\plotone{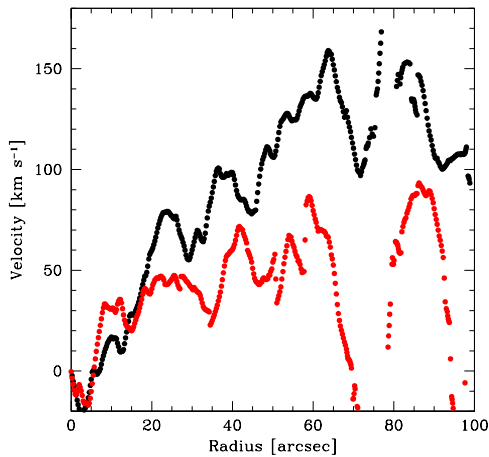}
\caption{Rotation curve determined from H$\alpha$ observations shown for each side of IC 2058, the only galaxy in the \cite{christlein} study that is also in the S$^4$G sample. }
\label{fig:rot}
\end{figure}

\cite{christlein} presented a study of 17 edge-on galaxies, in which they trace out the rotation of the galaxy using H$\alpha$ emission to as large a radius as possible. Unfortunately, of those galaxies only one, IC 2058, is also in the S$^4$G database. The rotation curve asymmetry is shown in Figure \ref{fig:rot} and is several tens of km s$^{-1}$ outside of 20 arcsec. Equation 2 suggests that the non-circular velocities will behave with radius as $(A_1/A_0)/R$. In general, Figure \ref{fig:profiles}, $A_1/A_0$ rises more slowly than $R$, suggesting that the non-circular velocity component should decline with radius. We are not seeing that in IC 2058, which instead shows a mild increase with radius, although it may be possible that in this galaxy $A_1/A_0$ is rising faster than $R$. Unfortunately, the spectroscopic data were not taken with care in defining the galaxy center, and the continuum of the galaxy is not sufficiently bright to allow an unambiguous center position. As such, it is not possible to pursue a more quantitative comparison using this spectrum and the S$^4$G data. 
Testing this correspondence in detail with a statistical sample would test the presumption of closed orbits in a non-axisymmetric potential. 

\section{Conclusions}

From a study of the $m=1$ and $m=2$ Fourier decompositions of the azimuthal 3.6$\mu$m surface brightnesses of 167 nearby galaxies covering a wide range of morphologies and luminosities, we confirm the previous claims of 1) a high (many tens of percent, depending on the choice of threshold) incidence on lopsidedness in the stellar distributions, 2) increasing lopsidedness as a function of radius over the stellar disk out to at least 3.5 scale lengths, and 3) larger lopsidedness, over these radii, for later type and lower surface brightness galaxies. 

In addition, we present the following new findings:

\noindent
The magnitude of the lopsidedness correlates with the character of the spiral arms. The stronger the arm pattern, ranging from the strongest, grand design, to multi-arm, then to the weakest, flocculent, the weaker the lopsidedness. This result is demonstrated quantitatively using the measures of the $m=1$ and $2$ distortions matched to visual arm classifications. We conclude that conditions that lead to lopsidedness tend to favor the genesis of flocculent arms, not strong, well defined ones.

\noindent
The magnitude of the lopsidedness has no detectable correlation with the presence or absence of a bar, or the strength of the bar when one is present. We interpret this finding to mean that the conditions that give rise to bar formation are unrelated to those generating most or all of the observed lopsidedness.

\noindent
Values of the stellar mass density, $f_*$, within a disk scale length are inversely correlated with lopsidedness. One interpretation we present is that this feature helps damp out $m=1$ modes both by anchoring the centers of the stellar and dark matter components and thereby minimizing ``sloshing" \citep{jm,kornreich}, and by circularizing any underlying asymmetries in the dark matter distribution because the stars oppose the potential asymmetry \citep{jog99, bailin}. An important caveat to this finding, and to a lesser extent in the next, is that we are unable to disentangle correlations with lopsidedness and $f_*$, from those with morphology and surface brightness.

\noindent
Values of $f_*$ over the interval range where lopsidedness is measured, 1.5 to 3.5 exponential scale lengths, correlate with values of lopsidedness. We hypothesize that when the stars are a significant fraction of the mass at those radii self-gravity can act effectively to generate or maintain the distortion.  

Together we interpret these findings as indicative of the following:

\noindent
Lopsidedness is a generic feature of galaxies and does not depend on a rare event, such as a direct accretion of a satellite galaxy onto the disk of the parent galaxy. Although such events must occasionally happen, and do give rise to features that match those observed \citep{walker,zaritsky}, they cannot explain the large incidence of lopsidedness nor all of the trends we observe without fine-tuning.

\noindent
Lopsidedness may be caused by several phenomena. Moderate lopsidedness ($\langle A_1 \rangle_i + \langle A_1 \rangle_o)/2 < 0.3$ reflects either halo asymmetries to which the disk responds or partially stabilized $m=1$ outer disk instabilities. Small asymmetries in the halo give rise to larger, observable stellar asymmetries \citep{jog}, and small asymmetries of a few percent in the halo are unavoidable and should be ubiquitous. The response by the stars to such an asymmetry depends both on how centrally concentrated they are with respect to the dark matter and whether there are enough stars in the region of the lopsidedness that self-gravity can become dynamically important.  The situation that favors stronger lopsidedness, relatively low stellar mass fractions at smaller radii are also likely to favor weak and irregular arms \citep[see][for a similar conclusion based on the analysis of rotation curves]{ee90}, tend to be found among later morphological types, and lead to lower overall surface densities \citep[which have been found to correlate with increased lopsidedness;][]{reichard09}. Alternatively, the high relative stellar mass densities found in the outer disks of lopsided galaxies are responsible for the instability that generates the disk morphology.

\noindent
Larger levels of lopsidedness are less dependent on the stellar mass fractions at small radii and probably arise more stochastically from major events in the history of the galaxy. Furthermore, lopsidedness in other components, namely the neutral gas, can have other origins, such as asymmetric gas accretion. 

The ubiquity of lopsidedness, the dynamical effects it must have on the disk and halo, the resulting influence on gas flows and star formation \citep{zaritsky,reichard09}, and the clues it can provide for aspects of galaxies that are difficult to study otherwise, such as the shape of their dark matter halos, are all reasons for the further study of this phenomenon.

\begin{acknowledgments}

DZ acknowledges financial support from 
NASA ADAP NNX12AE27G and thanks the Max Planck Institute for Astronomy and NYU CCPP for their hospitality during long-term visits. 
E.A. and A.B. acknowledge the CNES (Centre National d'Etudes Spatiales - France) for financial support. We acknowledge the support from the FP7 Marie Curie Actions of the European Commission, via the Initial Training Network DAGAL under REA grant agreement PITN-GA-2011-289313.
 The authors thank the entire  S$^4$G team for the efforts in making this program possible. This research has made use of the NASA/IPAC Extragalactic Database (NED), which is operated by the Jet Propulsion Laboratory, California Institute of Technology, under contract with NASA. This research 
is based in part on observations made with the Spitzer Space Telescope, which is operated by the Jet Propulsion Laboratory, California Institute of Technology under a contract with NASA. The National Radio Astronomy Observatory is a facility of the National Science Foundation operated under cooperative agreement by Associated Universities, Inc.
Taehyun Kim is a student at the National Radio Astronomy Observatory.
\end{acknowledgments}

\end{document}